\documentclass[11pt]{article}
\usepackage{amsfonts,amssymb, graphicx,a4,bm}

\newcommand{\qed}{\hfill\rule{3mm}{3mm}}
\newtheorem{teorema}{Theorem}

\newtheorem{proposition}[teorema]{Proposition}
\newtheorem{Remark}[teorema]{Remark}

\begin{document}


\voffset=-1.5truecm\hsize=16.5truecm    \vsize=24.truecm
\baselineskip=14pt plus0.1pt minus0.1pt \parindent=12pt
\lineskip=4pt\lineskiplimit=0.1pt      \parskip=0.1pt plus1pt

\def\ds{\displaystyle}\def\st{\scriptstyle}\def\sst{\scriptscriptstyle}


\let\a=\alpha \let\b=\beta \let\c=\chi \let\d=\delta \let\e=\varepsilon
\let\f=\varphi \let\g=\gamma \let\h=\eta    \let\k=\kappa \let\l=\lambda
\let\m=\mu \let\n=\nu \let\o=\omega    \let\p=\pi \let\ph=\varphi
\let\r=\rho \let\s=\sigma \let\t=\tau \let\th=\vartheta
\let\y=\upsilon \let\x=\xi \let\z=\zeta
\let\D=\Delta \let\F=\Phi \let\G=\Gamma \let\L=\Lambda \let\Th=\Theta
\let\O=\Omega \let\P=\Pi \let\Ps=\Psi \let\Si=\Sigma \let\X=\Xi
\let\Y=\Upsilon

\global\newcount\numsec\global\newcount\numfor
\gdef\profonditastruttura{\dp\strutbox}
\def\senondefinito#1{\expandafter\ifx\csname#1\endcsname\relax}
\def\SIA #1,#2,#3 {\senondefinito{#1#2}
\expandafter\xdef\csname #1#2\endcsname{#3} \else
\write16{???? il simbolo #2 e' gia' stato definito !!!!} \fi}
\def\etichetta(#1){(\veroparagrafo.\veraformula)
\SIA e,#1,(\veroparagrafo.\veraformula)
 \global\advance\numfor by 1
 \write16{ EQ \equ(#1) ha simbolo #1 }}
\def\etichettaa(#1){(A\veroparagrafo.\veraformula)
 \SIA e,#1,(A\veroparagrafo.\veraformula)
 \global\advance\numfor by 1\write16{ EQ \equ(#1) ha simbolo #1 }}
\def\BOZZA{\def\alato(##1){
 {\vtop to \profonditastruttura{\baselineskip
 \profonditastruttura\vss
 \rlap{\kern-\hsize\kern-1.2truecm{$\scriptstyle##1$}}}}}}
\def\alato(#1){}
\def\veroparagrafo{\number\numsec}\def\veraformula{\number\numfor}
\def\Eq(#1){\eqno{\etichetta(#1)\alato(#1)}}
\def\eq(#1){\etichetta(#1)\alato(#1)}
\def\Eqa(#1){\eqno{\etichettaa(#1)\alato(#1)}}
\def\eqa(#1){\etichettaa(#1)\alato(#1)}
\def\equ(#1){\senondefinito{e#1}$\clubsuit$#1\else\csname e#1\endcsname\fi}
\let\EQ=\Eq
\def\0{\emptyset}


\def\\{\noindent}
\let\io=\infty

\def\VU{{\mathbb{V}}}
\def\EE{{\mathbb{E}}}
\def\GI{{\mathbb{G}}}
\def\TT{{\mathbb{T}}}
\def\C{\mathbb{C}}
\def\CC{{\mathcal C}}
\def\II{{\mathcal I}}
\def\LL{{\cal L}}
\def\RR{{\cal R}}
\def\SS{{\cal S}}
\def\NN{{\cal N}}
\def\HH{{\cal H}}
\def\GG{{\cal G}}
\def\PP{{\cal P}}
\def\AA{{\cal A}}
\def\BB{{\cal B}}
\def\FF{{\cal F}}
\def\v{\vskip.1cm}
\def\vv{\vskip.2cm}
\def\gt{{\tilde\g}}
\def\E{{\mathcal E} }
\def\I{{\rm I}}

\def\tende#1{\vtop{\ialign{##\crcr\rightarrowfill\crcr
              \noalign{\kern-1pt\nointerlineskip}
              \hskip3.pt${\scriptstyle #1}$\hskip3.pt\crcr}}}
\def\otto{{\kern-1.truept\leftarrow\kern-5.truept\to\kern-1.truept}}
\def\arm{{}}
\font\bigfnt=cmbx10 scaled\magstep1

\newcommand{\card}[1]{\left|#1\right|}
\newcommand{\und}[1]{\underline{#1}}
\def\1{\rlap{\mbox{\small\rm 1}}\kern.15em 1}
\def\ind#1{\1_{\{#1\}}}
\def\bydef{:=}
\def\defby{=:}
\def\buildd#1#2{\mathrel{\mathop{\kern 0pt#1}\limits_{#2}}}
\def\card#1{\left|#1\right|}
\def\proof{\noindent{\bf Proof. }}
\def\qed{ \square}
\def\trp{\mathbb{T}}
\def\trt{\mathcal{T}}
\def\Z{\mathbb{Z}}
\def\be{\begin{equation}}
\def\ee{\end{equation}}


\title {Abstract polymer models with general pair interactions}

\author{ Aldo Procacci \\\footnotesize{Dep. Matem\'atica-ICEx, UFMG, CP 702
Belo Horizonte - MG, 30161-970 Brazil}\\\footnotesize{ email:~ aldo@mat.ufmg.br}}
\date{}
\maketitle
\begin{abstract}
A convergence criterion of cluster expansion is presented in the
case of an abstract polymer system with general pair interactions
(i.e. not necessarily hard core or repulsive).
As a concrete example,  the
low temperature disordered phase of the BEG model with infinite range interactions,
decaying polynomially as  $1/r^{d+\l}$ with $\l>0$, is studied.
\end{abstract}
\vv\vv\vv
\numsec=1\numfor=1
\\{\bf\Large 1. Introduction}
\vv
The abstract polymer gas  is an important tool to study the
high temperature/low density or low temperature phase of many
statistical mechanics models.  Generally speaking, the abstract polymer model
consists of a collection of objects (the polymers)  which play the
role of the particles of the gas. These polymers have a given
activity and they interact via a hard core pair potential
suitably defined. Typically, one wants to show that the pressure
of this polymer gas can be written in terms of an
absolutely convergent series if the activities are
taken sufficiently small.

The first example of such a model appeared in \cite{GK} where the
polymers were finite non overlapping subsets of the cubic lattice
$\Z^d$. The authors proved convergence of the pressure  via the
method of Kirkwood-Salsburg equations. Subsequently,  the same
system studied in \cite{GK} was treated in \cite{Se} and \cite{C}
via cluster expansion methods  based on tree graph inequalities.

In \cite{KP} the most general version of this system was given. There, polymers were simply a collection of objects
with a given activity and interacting through an hard core pair potential introduced via a symmetric and reflexive
relation in the polymer space. Polymers belonging to this relation  were called incompatible,
and compatible otherwise. The hard core condition was  simply to forbid configurations of polymers
containing pairs of incompatible polymers.
Differently from the cases considered previously,  in which polymers had a
cardinality and a size, the Koteck\'y-Preiss polymers were characterized only by the activity.

In \cite{D} the convergence condition for the Koteck\'y-Preiss polymer gas was slightly improved and the proof
was greatly simplified, being reduced to a simple inductive argument, as it was shown very clearly
in \cite{Mi} and \cite{S1}. In particular,
in \cite{S1} it has been observed  that the Dobrsushin's proof  works  even  for
more general abstract polymer gases, in which polymers
may interact through a repulsive soft-core pair interaction.

Very recently \cite{FP} the Koteck\'y-Preiss and the Dobrushin conditions for convergence
of the abstract polymer gas with purely hard core interactions were reobtained via the standard cluster expansion methods and
a new improved condition was given by exploiting   an old tree graph identity valid for hard core systems
due to O. Penrose \cite{Pe}.

\newpage
In all these works, the basically hard core character of the interaction seemed to be an essential
ingredient to control the convergence. Exceptions  can be found in \cite{DS}, \cite{I}.
In \cite{DS} a contour model with interaction (exponentially decaying al large distances)
is proposed. However the model is rewritten in term
of the usual hard core polymer gas where polymers are objects more complicated than the original contours.
This philosophy has also been pursued in \cite{I} where a one-dimensional contour model
with long range interaction is rewritten in term of new objects with hard core pair interactions.

It would be of interest to treat also cases in which polymers interact via more general pair
interactions, e.g., not necessarily repulsive, not necessarily hard core, not necessarily finite range.
Such abstract polymer model  could be a useful tool  in the study of spin systems at low temperature
interacting via infinite range polynomially decaying
potential, see e.g. \cite{Pa}.

In this paper we develop a model of abstract polymers (of the type of \cite{KP}) with  interactions
more general than the hard-core. Our polymers interact through a "short distance" repulsive (not necessarily hard core)
pair potential which is non zero only on pair of incompatible polymers, plus an
a pair potential with no definite sign (hence it can be attractive), acting only on pairs of compatible polymers.
We give a condition convergence for the pressure of this gas by using a cluster expansion method
similar to the one developed in \cite{FP}. However, differently from \cite{FP}, we could not use here
the Penrose identity, since our interaction is not purely hard-core. We rather used another well known
tree graph identity originally proposed in \cite{BrF} and further developed in \cite{BaF}.

The rest of the paper is organized as follows. In section 2 we introduce the model, notations and the main result
of the paper. In section 3 we give the proof of our result (theorem 1).
Namely, in subsection 3.1. we present the tree graph identity and show how it can be used to bound the Ursell
coefficients of the Mayer series of our polymer model. In  subsection 3.2 we give
the  convergence argument based on map iterations developed in \cite{FP}. In subsection 3.3 we conclude the proof of our main theorem.
Finally in section 4, as an example, we use theorem 1 to study  the
low temperature disordered phase of the BEG model with infinite range interactions
with polynomial decay of the type $1/r^{d+\l}$ with $\l>0$.

\numsec=2\numfor=1
\vv\vv\vv
\\{\bf\Large 2. Polymer gas: notations and  results}
\vv

\\{\bf 2.1. The model}.
\vv
\\Let $\PP$ denotes the set of polymers (i.e. $\PP$ is the
{\it single particle state space}). We will assume here that $\PP$ is a countable set.
We associate to each polymer $\g\in \PP$ a complex  number $z_\g$ (a positive number in physical situations)
which is  interpreted as the  {\it activity} of the  polymer $\g$. We will denote $z=\{z_\g\}_{\g\in \PP}$.

Polymers interact through a pair potential. Namely, the energy $E$ of a configuration $\g_1,\dots,\g_n$
of $n$ polymers is given by
$$
E(\g_1,\dots,\g_n)= \sum_ {1\leq i<j\leq n}V(\g_i,\g_j)\Eq(U)
$$
where pair potential $V(\g,\g')$ is a symmetric function in $\PP\times \PP$ taking values in
$ \mathbb{R}\cup\{+\infty\}$. Observe that we don't make any hypothesis on the sign of $V(\g_i ,\g_j)$
so this interaction could
be for some pairs attractive and for other pairs  repulsive.

Fix now  a finite set $\L\subset \PP$ (the "volume" of the gas). Then the probability to see the configuration
$(\g_{1},\dots ,\g_{n})\in \L^n$ is given by
$$
Prob(\g_{1},\dots ,\g_{n})={1\over \Xi_{\L}} {{z_{\g_1}}{z_{\g_2}}\dots{z_{\g_n}}
e^{- \sum_ {1\leq i<j\leq n}V(\g_i,\g_j)}}
$$
where the normalization constant $\Xi_{\L}$ is
the grand-canonical partition function  in the volume $\L$ and  is given by
$$
\Xi_{\L}(z)=1+\sum_{n\geq 1}{1\over n!}
\sum_{(\g_{1},\dots ,\g_{n})\subset\L^n}
{z_{\g_1}}{z_{\g_2}}\dots{z_{\g_n}}
e^{- \sum_ {1\leq i<j\leq n}V(\g_i,\g_j)}\Eq(partiz)
$$
Note that the configurations $\g_1,\dots,\g_n$ for there exist a pair $\g_i,\g_j$ such that $V(\g_i,\g_j)=+\infty$
have zero probability to occur, i.e. are  forbidden. So, following the tradition, if a pair $(\g,\g')\in \PP\times \PP$
is
such that $V(\g,\g')=+\infty$, we will denote by $\g\not\sim\g'$ and  say that $\g$ and $\g'$ are  {\it incompatible}.

Since we are admitting non purely repulsive interaction among polymers,
we  also need to require that the potential energy $E$ is stable in the classical sense. This can be achieved
by imposing that
there exists a function $B(\g)\ge 0$ such that
$$
\sum_ {1\leq i<j\leq n}V(\g_i,\g_j)\ge -\sum_{i=1}^nB(\g_i)\Eq(stab)
$$
for all $n\in \mathbb{N}$ and all $(\g_1, \dots, \g_n)\in \PP^n$. Note that in the case in which $(\g_1, \dots, \g_n)$
contains pairs of incompatible polymers the l.h.s. of \equ(stab) is $+\infty$ so this inequelity is trivially satisfied.

The stabiltity condition immediately
implies that $\X_\L$ is convergent and
$$
\X_\L\le 1+\sum_{n\geq 1}{1\over n!}
\Bigg[\sum_{\g\subset\L}
z_{\g}
e^{B(\g)}\Bigg]^n\le \exp\Big\{\sum_{\g\in \L}z_\g e^{B(\g)}\Big\}\le |\L|\max_{\g\in \L}\exp\{z_\g e^{B(\g)}\}
$$
Actually, \equ(stab) implies that $\X_\L(z)$ is  analytic in the whole $\mathbb{C}^{|\L|}$ ($|\L|$ is the cardinality of $\L$).

As we said in the introduction, the usual choice available in the literature is that
$V(\g,\g')$ takes values in the set $\{0,+\infty\}$ for all $(\g,\g')\in\PP\times\PP$ and
$V(\g,\g)=+\infty$  for all $\g\in \PP$ (purely hard core pair potential) but 
we remark that the purely repulsive case 
(i.e. $0\le V(\g_i, \g_j)\le +\infty$  for all pairs) has also been  considered in 
\cite{S1} and \cite{U}.
However, in view of the possible connections with the low temperature phase
of spin systems with infinite range interactions,
we think that the most interesting situation treated in the
present paper is the case $V(\g_i ,\g_j)< 0$, i.e. when an attractive potential, possibly
infinite range, is acting among polymers.

\vv\vv
\\{\bf 2.2. Results}.
\vv
\\The  pressure of this gas, namely $\log \Xi_{\L}$, can
be written as a formal series through a
Mayer expansion
on the Gibbs factor
$\exp\{-\sum_ {1\leq i<j\leq n}V(\g_i,\g_j)\}$.
Namely, a standard calculations (see e.g. \cite{C}) gives
$$
\log \Xi_{\L}(z)= \sum_{n=1}^{\infty}{1\over n!}
\sum_{(\g_{1},\dots ,\g_{n})\subset\L^n}
\phi^{T}(\g_1 ,\dots , \g_n)z_{\g_1}\dots z_{\g_n}\Eq(6)
$$
with
$$
\phi^{T}(\g_{1},\dots ,\g_{n})=\cases{1&if $n=1$\cr\cr
\sum\limits_{g\in G_{n}}\prod\limits_{\{i,j\}\in E_g}(e^{-V(\g_i,\g_j)}-1)&if $n\ge 2$
}
\Eq(7)
$$
where  ${G}_n$  is the set all connected
graphs with vertex set $\{1,2,\dots,n\}$. We recall that
a graph $g\in G_n$ is a pair
$g=(V_g,E_g)$ where $V_g=\{1,2,\dots,n\}$ is the set of vertices of $g$ and
$E_g\subset \{\{i,j\}\subset \{1,2,\dots ,n\}\} $ is the set of edges of $g$.  We also recall that $g=(V_g,E_g)$ is connected if
for any $A,B$ such that $A\cup B= V_g$ and $A\cap B=\0$, there exists $\{i,j\}\in E_g$ such
That $A\cap\{i,j\}\neq \0$ and $B\cap\{i,j\}\neq \0$.

The equation \equ(6) makes sense  only for those $z$ for which  the formal series in the r.h.s. of \equ(6) converges absolutely.
To study  absolute convergence,
we will consider the positive term series
$$
|\log \Xi_{\L}|(\r)=
\sum_{n=1}^{\infty}{1\over n!}
\sum_{(\g_{1},\dots ,\g_{n})\subset\L^n}
|\phi^{T}(\g_1 ,\dots , \g_n)|{\r_{\g_1}}\cdots{\r_{\g_n}}
\Eq(6abs)
$$
where now $\r_\g\in [0,+\infty)$, for all $\g\in \PP$ and $\r=\{\r_\g\}_{\g\in \PP}$. Of course $|\log \Xi_{\L}(z)| \le  |\log \Xi_{\L}|(\r)$
for $z$ in the polydisk $\{|z_\g|\le \r_\g\}_{\g\in \PP}$.

We further define, for each $\g_0\in \PP$,  a  function $\Pi^{\g_0}_{\PP}(\r)$
directly related to \equ(6abs) (a ``pinned'' sum defined in the whole set $\PP$) as follows

$$
\Pi^{\g_0}_{\PP}(\r)=\sum_{n=0}^{\infty}{1\over n!} \sum_{(\g_1,\g_2,\dots,\g_n)\in \PP^n}
|\phi^{T}(\g_0,\g_1 ,\dots , \g_n)|{\r_{\g_1}}\dots{\r_{\g_n}}\Eq(TP)
$$

\\Clearly, if we are able to show that $\Pi^{\g_0}_{\PP}(\r)$ converges, then
$|\log \Xi_{\L}|(\r)$ and hence $|\log \Xi_{\L}(z)|$ for $|z_\g|\le \r_\g$  also converge, since
it is easy to check that
$$
 |\log \Xi_{\L}|(\r)\le
|\L|\sup_{\g_0\in \L}\r_{\g_0}\Pi^{\g_0}_{\PP}(\r)\Eq(PminTP)
$$

To understand the meaning of the series $\Pi^{\g_0}_{\PP}(z)$  just observe that its
finite volume version  $\Pi^{\g_0}_{\PP}(\r)$,
namely
$$
\Pi^{\g_0}_{\L}(\r)=\sum_{n=0}^{\infty}{1\over n!} \sum_{(\g_1,\g_2,\dots,\g_n)\in \L^n}
|\phi^{T}(\g_0,\g_1 ,\dots , \g_n)|{\r_{\g_1}}\dots{\r_{\g_n}}\Eq(PiLa)
$$
is directly related to $\log \Xi_{\L}(\r_\L)$. It is immediate  to see that

$$
\Pi^{\g_0}_{\L}(\r)={\partial\over \partial \r_{\g_0}}|\log \Xi_{\L}|(\r)
\Eq(relpilog)
$$

\noindent

The main result of the paper is
a convergence criterion for the positive series \equ(TP). Such criterion can be
considered as a generalization of the Koteck\'y-Preiss criterion for polymer
system interacting through a pair potential which is not purely hard core. The criterion can be stated
as the following theorem.

\begin{teorema}\label{th:1}.
Let $\m: \PP\to [0,\infty): \g\mapsto \m_\g$ be a non negative valued function and let, for each $\g\in \PP$, $\rho_\g\in [0,\infty)$
such that
$$
\r_\g e^{B(\g)}\le {\m_\g}\,e^{-\sum_{\tilde\g\in\PP} F(\g,\tilde\g)
{\mu_{\tilde\g}}}, \;\;\;\,\,\,\,\,\,\,\forall \g\in \PP \Eq(muRv)
$$
where $B(\g)$ is the function defined in \equ(stab) and
$$
F(\g_i,\g_j)=\cases{
\left|e^{-V(\g_i,\g_j)}-1\right|=1 &if $\g_i\nsim \g_j$
\cr\cr
|V(\g_i,\g_j)| &otherwise
}\Eq(defF)
$$

\\Then the series $\Pi_{\g_0}(\r)$ [defined in \equ(TP)] converges and satisfies
$\r_{\g_0}\,\Pi_{\g_0}(\r)\le\mu_{\g_0}$.
\end{teorema}

{\bf Remark}. Observe that in the usual case $U$ hard-core  one obtains  from theorem \ref{th:1}
the usual Koteck\'y-Preiss condition.
We recall however when polymers interact just through a purely repulsive potential, one can do better
than \equ(muRv). In particular,
for the  purely hard core case
it has been shown in \cite{FP}
that the condition \equ(muRv) can be considerably improved by taking advantage of the Penrose tree identity \cite{Pe},
(see also \cite{Pf}, \cite{S1} \cite{FP}) valid in the case of purely hard core interactions.

\numsec=3\numfor=1
\vv\vv\vv
\\{\Large \bf 3. Proof of theorem 1}.
\vv

\vv\vv

\\The strategy of the proof is quite similar to the one used in \cite{FP}.  In particular
we use here the very same convergence argument for positive series which has been developed in \cite{FP}.
On the other hand, in the present case we cannot use the Penrose identity in order to bound the Ursell
coefficients $|\phi^{T}(\g_1 ,\dots , \g_n)|$, since the pair potential  is not purely hard-core
(and also not purely repulsive).
We will rather make
use of another well known ``tree graph identity'' originally
proved in \cite{BrF} (see also \cite{Br, BaF, PdLS, Book}).

\vv\vv
\\{\bf 3.1. Tree graph  inequality  for $|\phi^{T}(\g_0,\g_1 ,\dots , \g_n)|$}
\vv
\\We state the so called tree graph identity \cite{BrF},\cite{BaF}, \cite{Br} by
using the notations of \cite{PdLS} and \cite{Book}. We use the short notation $\I_n=\{1,2,\dots ,n\}$.  A graph $\t=(\I_n,E_\t)\in G_n$ is called
a {\it tree}
if and only if its edge set $E_\t$ has cardinality equal to $n-1$.
Let us denote by $T_n$ the set of trees
with vertex set $\I_n$.

In the following whenever $U$ is a finite set, $|U|$ denotes its cardinality.

\vskip.3cm
\\ {\bf Lemma 2 (Tree graph identity)}. {\it
Let $V_{ij}$, with $1\le i<j\le n $ be $n(n-1)/2$ real numbers,
then the following identity holds
$$
\sum_{g\in G_n}
\prod_{\{i,j\}\in E_g}\left(e^{- V_{ij}}-1\right)=
\sum_{\t\in T_n}\prod_{\{i,j\}\in E_\t} (- V_{ij})
\int d\m_{\t}({\bf t}_{n-1},{\bf X}_{n-1})
e^{-K({\bf X}_{n-1},{\bf t}_{n-1})}\Eq(TGI)
$$

\\where:
\begin{itemize}
\item
${\bf t}_{n-1}$ denote a set on $n-1$ interpolating parameters
${\bf t}_{n-1}\equiv (t_1 ,\dots ,t_{n-1})\in [0,1]^{n-1}$;
\item
${\bf X}_{n-1}$ denote a set of ``increasing'' sequences of $n-1$ subsets,
${\bf X}_{n-1}\equiv X_1 ,\dots ,X_{n-1}$ such as
$\forall i$,  $X_{i}\subset \I_n$, we must have
$X_{i}\subset X_{i+1}$, $|X_{i}|=i$ and $X_1 =\{1\}$.

\item
$K({\bf X}_{n-1},{\bf t}_{n-1})$ is  a convex decomposition of the potential,
explicitly given by
$$
K({\bf X}_{n-1},{\bf t}_{n-1}) =  \sum_{1\leq i<j\leq n}
t_{1}(\{i,j\})\dots t_{n-1}(\{i,j\})V_{ij}\Eq(convex)
$$
where
$$
t_{l}(\{i,j\})=\cases{ t_{l}\in [0,1]
&if
$i\in X_l$ and $j\notin X_l$ or vice versa\cr
1
&otherwise\cr}
$$
(a pair $\{i,j\}$  such that
$i\in X_l$ and $j\notin X_l$ or vice versa is said to ``cross''
$X_l$).

\item
The measure

$$
\int d\m_{\t}({\bf t}_{n-1},{\bf X}_{n-1})\,\,[\cdots]
\doteq
\int_{0}^{1}dt_1\dots\int_{0}^{1}dt_{n-1}
\sum_{{\bf X}_{n-1}\atop {\rm comp.}\ \t}
t_{1}^{b_{1}-1}\dots
t_{n-1}^{b_{n-1}-1}\,\,[\cdots]\Eq(measure)
$$
has total mass equal to one (i.e. it is a probability measure).
In \equ(measure)
"${\bf X}_{n-1}
{\rm \,comp.}\,\t$" means that
for all $i =1,2,\dots ,n-1$,
$X_i$ contains exactly $i-1$ edges of $\t$ and
$b_i$ is the numebr of edges  in $\t$
which ``cross'' $X_i$
\end{itemize}
}
\vskip.2cm

We want to use  \equ(TGI) to bond $|\phi^{T}(\g_1 ,\dots , \g_n)|$. This formula is  useful
when the pair potential is not purely repulsive. However, due to the restriction $V_{ij}$ finite (otherwise the r.h.s. of \equ(TGI)
is not well defined),
one in general can apply \equ(TGI) only if  the pair potential is finite and absolutely integrable, see \cite{Br},
which is a quite restrictive condition. In  particular
this rules out a pair potential with hard core at short distances which is precisely one of the situations we would like to treat.

We show here that it is possible to give meaning to r.h.s. of \equ(TGI) even
when some among the $V_{ij}$'s take the value $\infty$ (the l.h.s. of \equ(TGI) makes sense even in this case).  We define
a cut-offed pair potential
$$
V_{H}(\g_i,\g_j)=\cases{ H & if $\g_i\nsim\g_j$\cr\cr
V(\g_i,\g_j) &otherwise
}\Eq(VH)
$$
Note that, from stability condition \equ(stab), for any fixed $n\in \mathbb{N}$ and  $(\g_1,\dots,\g_n)\in \PP^n$,
there is $H_0$ (which depends on $n$ and  $(\g_1,\dots,\g_n)\in \PP^n$) such that, for all $H\ge H_0$ and for all $X\subset \{1,2,\dots,n\}$,
$$
\sum_ {\{i,j\}\subset X}V_H(\g_i,\g_j)\ge -\sum_{i\in X}B(\g_i)
\Eq(stabH)
$$
Indeed,
if $X\subset \{1,2,\dots,n\}$ is such that $\{\g_i\}_{i\in X}$ does not contain incompatible pairs, then, by definition \equ(VH) and
inequality \equ(stab), it follows
$$
\sum_ {\{i,j\}\subset X}V_H(\g_i,\g_j)=
\sum_ {\{i,j\}\subset X}V(\g_i,\g_j)\ge \sum_{i\in X}B(\g_i)
$$
If $X$ is such that $\{\g_i\}_{i\in X}$ does contain incompatible pairs,
then there is at least an edge $\{k,s\}$ such that $V^H(\g_i,\g_j)=H$ so taking
$$
H_0^X= -\sum_{\{i,j\}\in X\atop V(\g_i,\g_j)\le 0,\,\, \{i,j\}\neq \{k,s\}}V(\g_i,\g_j)
$$
we have, whenever $H\ge H_0^X$
$$
\sum_ {\{i,j\}\subset X}V_H(\g_i,\g_j)\ge\; 0\;\ge  -\sum_{i\in X}^nB(\g_i)
$$
So, taking $H_0=\max_{X\subset \{1,2,\dots,n\}} H_0^X$ the inequalities \equ(stabH) are satisfied for all
$X\subset \{1,2,\dots,n\}$.

\vv\vv
\\Now, for any fixed
$(\g_1, \dots , \g_n)\in \PP^n$
$$
\phi^{T}(\g_1 ,\dots , \g_n)=\lim_{H\to \infty}
\sum\limits_{g\in G_{n}}\prod\limits_{\{i,j\}\in E_g}(e^{- V_{H}(\g_i,\g_j)}-1)
$$
We can now use \equ(TGI) for the {\it finite} potential $V_H$ and we get
$$
\sum\limits_{g\in G_{n}}\prod\limits_{\{i,j\}\in E_g}(e^{- V_{H}(\g_i,\g_j)}-1)=
$$
$$=
\lim_{H\to\infty}
\sum_{\t\in T_n}\prod_{\{i,j\}\in E_\t} (- V_{H}(\g_i,\g_j))
\int d\m_{\t}({\bf t}_{n-1},{\bf X}_{n-1})
e^{- K_{H}({\bf X}_{n-1},{\bf t}_{n-1})}
$$
where
$$
K_{H}({\bf X}_{n-1},{\bf t}_{n-1}) =  \sum_{1\leq i<j\leq n}
t_{1}(\{i,j\})\dots t_{n-1}(\{i,j\})V_{H}(\g_i,\g_j)\Eq(KH)
$$

Now, for fixed $\t=(\I_n,E_\t)\in T_n$  and $(\g_1, \dots, \g_n)\in \PP^n$ , let us consider the  factor
$$
w^\t_H(\g_1, \dots, \g_n)= \prod_{\{i,j\}\in E_\t} |V_{H,J}(\g_i,\g_j)|
\int d\m_{\t}({\bf t}_{n-1},{\bf X}_{n-1})
e^{-  K_{H}({\bf X}_{n-1},{\bf t}_{n-1})}
$$
\def\E{{\rm E}}

\\The edges $\{i,j\}\subset \I_n$ are naturally partitioned into two disjoint sets
$\E_n^H$ and $\E_n\backslash \E^H_n$ where
$\E_n^H=\{\{i,j\}\subset I_n : \g_i\not\sim \g_j)\}$.
Thus also the edges
of the tree $\t$ are partitioned  into two disjoint sets
$E_\t^H$ and $E_\t\backslash E_\t^H$ where
$E_\t^H=E_\t\cap \E^H_n$. So we have
$$
w^\t_H(\g_1, \dots, \g_n)
=  \prod_{\{i,j\}\in E_\t\backslash E_\t^H}  |V_H(\g_i,\g_j)|~\prod_{\{i,j\}\in  E_\t^H} |V_H(\g_i,\g_j)|
\int d\m_{\t}({\bf t}_{n-1},{\bf X}_{n-1})
e^{-  K_{H}({\bf X}_{n-1},{\bf t}_{n-1})}\Eq(wtaui)
$$
Now, recalling the definition \equ(KH), we can write
$$
K_{H}({\bf X}_{n-1},{\bf t}_{n-1})= K_{U_{(1-\e)H}}({\bf X}_{n-1},{\bf t}_{n-1})+
K_{V_{\e H}} ({\bf X}_{n-1},{\bf t}_{n-1})
$$
where
$$
K_{U_{(1-\e)H}}({\bf X}_{n-1},{\bf t}_{n-1})=
\sum_{1\leq i<j\leq n}
t_{1}(\{i,j\})\dots t_{n-1}(\{i,j\})U_{{(1-\e)H}}(\g_i,\g_j)
$$
and
$$
K_{V_{\e H}}({\bf X}_{n-1},{\bf t}_{n-1})=
\sum_{1\leq i<j\leq n}
t_{1}(\{i,j\})\dots t_{n-1}(\{i,j\})V_{\e H}(\g_i,\g_j)
$$
where $\e>0$ and
$$
U_{(1-\e)H}(\g_i,\g_j)=\cases{ (1-\e)H & if $\g_i\nsim\g_j$\cr\cr
0 &otherwise
}
$$
and
$$
V_{\e H}(\g_i,\g_j)=\cases{ \e H & if $\g_i\nsim\g_j$\cr\cr
V(\g_i,\g_j) &otherwise
}
$$
The potential $t_{1}(\{i,j\})\dots t_{n-1}(\{i,j\})V_{\e H}(\g_i,\g_j)$ satisfies,
for $H$ larger that $\e^{-1}H_0$
$$
\sum_ {\{i,j\}\subset X}V_{\e H}(\g_i,\g_j)\ge -\sum_{i\in X}B(\g_i)
$$
for all $X\subset \{1,2,\dots,n\}$. This fact implies (see e.g. \cite{Br}, \cite{Book}, \cite{PdLS}) that
$$
K_{V_{\e H}}({\bf X}_{n-1},{\bf t}_{n-1})\ge
-\sum_{i=1}^n B(\g_i) \Eq(kcomp)
$$

\\The potential $K_{U_{(1-\e)H}}({\bf X}_{n-1},{\bf t}_{n-1})$ is non negative and
can be bounded, for $\h>0$, as follows
$$
K_{U_{(1-\e)H}}({\bf X}_{n-1},{\bf t}_{n-1})
\ge
\sum_{\{i,j\}\subset E_\t^H}
t_{1}(\{i,j\})\dots t_{n-1}(\{i,j\})U_{(1-\e)H}(\g_i,\g_j)=
$$
$$
=
\sum_{\{i,j\}\subset E_\t^H}
t_{1}(\{i,j\})\dots t_{n-1}(\{i,j\})(1-\e)H
+
\h\sum_{\{i,j\}\subset E_\t\backslash E_\t^H}
t_{1}(\{i,j\})\dots t_{n-1}(\{i,j\}) -
$$
$$-\h
\sum_{\{i,j\}\subset E_\t\backslash E_\t^H}
t_{1}(\{i,j\})\dots t_{n-1}(\{i,j\})~~\ge
$$
$$
\ge
\sum_{\{i,j\}\subset E_\t^H}
t_{1}(\{i,j\})\dots t_{n-1}(\{i,j\})(1-\e)H+\!\!\!\!\!
\sum_{\{i,j\}\subset E_\t\backslash E_\t^H}
t_{1}(\{i,j\})\dots t_{n-1}(\{i,j\})\h  - {|E_\t\backslash E_\t^H|\h}
$$
So we get
$$
K_{U_{(1-\e)H}}({\bf X}_{n-1},{\bf t}_{n-1})\ge \sum_{1\leq i<j\leq n}
t_{1}(\{i,j\})\dots t_{n-1}(\{i,j\})V^\t_{ij} - |E_\t\backslash E_\t^H| \eta\Eq(kin)
$$
where
$V^\t_{ij}$ is the positive ($H,\h,\e$ dependent) pair potential given by
$$
V^\t_{ij}=\cases{(1-\e)H &if $\{i,j\}\in E_\t^H$ \cr
{\h} & if $\{i,j\}\in E_\t\backslash E_\t^H$\cr
0 &otherwise
}
$$

\\Hence, plugging \equ(kcomp) and \equ(kin) into \equ(wtaui) we obtain that $w^\t_H(\g_1, \dots, \g_n)$ can be bounded by
$$
w^\t_H(\g_1, \dots, \g_n)\le
e^{ +\sum_{i=1}^n B(\g_i)+ \h|E_\t\backslash E_\t^H|}
\left[\prod_{\{i,j\}\in E_\t\backslash E_\t^H} |V(\g_i,\g_j)|\right]
\times\left[{1\over \h}\right]^{|E_\t\backslash E_\t^H|}\times
$$
$$
\times \left[{1\over 1-\e}\right]^{|E_\t^H|}
\prod_{\{i,j\}\in  E_\t}  V^\t_{ij}
 \int d\m_{\t}({\bf t}_{n-1},{\bf X}_{n-1})
e^{-  \sum_{1\leq i<j\leq n}
t_{1}(\{i,j\})\dots t_{n-1}(\{i,j\})V^\t_{ij} }
$$
Applying now the tree graph identity \equ(TGI) to the pair potential $V^\t_{ij}$
one conclude immediately (see e.g. \cite{Br}) that, for all $H\in [0,+\infty)$
$$
\prod_{\{i,j\}\in  E_\t}  V^\t_{ij}
 \int d\m_{\t}({\bf t}_{n-1},{\bf X}_{n-1})
e^{-  \sum_{1\leq i<j\leq n}
t_{1}(\{i,j\})\dots t_{n-1}(\{i,j\})V^\t_{ij} }=\prod_{\{i,j\}\in E_\t}\left|e^{-V^\t_{ij}}-1\right|=
\Eq(bry)
$$
$$
=
\left|e^{-(1-\e)H}-1\right|^{|E_\t^H|}
\left|e^{-\h}-1\right|^{|E_\t\backslash E_\t^H|}
=
\left|e^{-\h}-1\right|^{|E_\t\backslash E_\t^H|} \prod_{\{i,j\}\in E_\t^H}\left|e^{-U_{(1-\e)H}(\g_i,\g_j)}-1\right|
$$

\\Hence, considering that
$e^{ \h|E_\t\backslash E_\t^H|}\left|e^{-\h}-1\right|^{|E_\t\backslash E_\t^H|}=(e^\h-1)^{ |E_\t\backslash E_\t^H|}$
we get
$$
w^\t_H(\g_1, \dots, \g_n)\le
e^{ +\sum_{i=1}^n B(\g_i)}\prod_{\{i,j\}\in E_\t^H} \left[{1\over 1-\e}\right]\left|e^{-U_{(1-\e)H}(\g_i,\g_j)}-1\right|
\prod_{\{i,j\}\in E_\t\backslash E_\t^H} \Bigg|{(e^\h-1)\over \h}V(\g_i,\g_j)\Bigg|
$$
and due to the arbitrarity of $\h$ and $\e$ which can be taken as small as we please, and using also
that $U_{(1-\e)H}(\g_i,\g_j)<V(\g_i,\g_j)$ for any $\g_i\not\sim\g_j$ and any finite $H$,   we obtain, for any $H\ge H_0$
$$
w^\t_H(\g_1, \dots, \g_n)\le
e^{ +\sum_{i=1}^n B(\g_i)} \prod_{\{i,j\}\in E_\t^H}\left|e^{-V(\g_i,\g_j)}-1\right|
\prod_{\{i,j\}\in E_\t\backslash E_\t^H}  |V(\g_i,\g_j)|
$$
which is a bound independent of $H$. So

$$
w^\t(\g_1, \dots, \g_n)=\lim_{H\to \infty} w^\t_H(\g_1, \dots, \g_n)\le
e^{ +\sum_{i=1}^n B(\g_i)} \prod_{\{i,j\}\in E_\t^H}\left|e^{-V(\g_i,\g_j)}-1\right|
\prod_{\{i,j\}\in E_\t\backslash E_\t^H}  |V(\g_i,\g_j)|
$$
In conclusion we have that
$$
|\phi^{T}(\g_1, \dots, \g_n)|\le  e^{ +\sum_{i=1}^n B(\g_i)}
\sum_{\t\in T_n} \prod_{\{i,j\}\in E_\t}  F(\g_i,\g_j)\Eq(burs)
$$
where
$$
F(\g_i,\g_j)=\cases{
\left|e^{-V(\g_i,\g_j)}-1\right| & if $\g_i\nsim \g_j$ \cr\cr
|V(\g_i,\g_j)| &otherwise
}
$$
and hence also, for $n\ge 1$
$$
|\phi^{T}(\g_0,\g_1, \dots, \g_n)|\le  e^{ +\sum_{i=0}^n B(\g_i)}
\sum_{\t\in T^0_{n}} \prod_{\{i,j\}\in E_\t}  F(\g_i,\g_j)\Eq(bursa)
$$
where $T^0_n$ is the set of all trees with vertex set $\I_n^0\doteq\{0,1,2,\dots, n\}$. Inserting \equ(bursa)
in \equ(TP) we get
$$
\Pi^{\g_0}_{\PP}(\r)\le 1+\sum_{n=1}^{\infty}{1\over n!} \sum_{(\g_1,\g_2,\dots,\g_n)\in \PP^n}
 e^{ +\sum_{i=0}^n B(\g_i)}
\sum_{\t\in T^0_{n}} \prod_{\{i,j\}\in E_\t}  F(\g_i,\g_j){\r_{\g_1}}\dots{\r_{\g_n}}\le
$$
$$
\le e^{B(\g_0)}\left[1+\sum_{n=1}^{\infty}{1\over n!} \sum_{(\g_1,\dots,\g_n)\in \PP^n}
\sum_{\t\in T^0_{n}} \prod_{\{i,j\}\in E_\t}  F(\g_i,\g_j)\r_{\g_1}e^{B(\g_1)}\dots\r_{\g_n}e^{B(\g_n)}\right]=
$$

\\If we pose

$$
\tilde\r_\g= \r_\g e^{B(\g)}\Eq(rhotil)
$$

\\and

$$
|\tilde\Pi|^{\g_0}_{\PP}(\tilde\r)= 1+\sum_{n=1}^{\infty}{1\over n!}
\sum_{\t\in T^0_{n}} \sum_{(\g_1,\dots,\g_n)\in \PP^n} \prod_{\{i,j\}\in E_\t}  F(\g_i,\g_j){\tilde\r_{\g_1}}\dots{\tilde\r_{\g_n}}\Eq(mukotpre)
$$

\\We get

$$
|\Pi|^{\g_0}_{\PP}(\r)\le e^{B(\g_0)} |\tilde\Pi|^{\g_0}_{\PP}(\tilde\r)\Eq(dises)
$$

\\So the convergence of $|\tilde\Pi|^{\g_0}_{\PP}(\tilde\r)$ implies that of $|\Pi|^{\g_0}_{\PP}(\r)$.

\vv\vv
\\{\bf 3.2.  Planar rooted trees and convergence}
\vv
\\We think the trees with vertex set $\I_n^0=\{0,1,2,\dots,n\}$ (i.e the elements of $T^0_n$) as rooted in $0$.
We define a map $m:\t\mapsto m(\t)$ which associate to each labelled tree $\t\in T^0_{n}$ a unique  drawing $t=m(\t)$ in the plane,
called the {\it planar rooted tree} associated to $\t$,
as follows.

Given $\t$ in $T^0_{n}$, place the vertex $0$ (the root)
at the leftmost position of the
drawing.  From 0 there emerge $s_{0}$ branches ending at the
{\it first-generation} vertices $i_1,\dots, i_{s_0}$. Drawn these vertices along a vertical line
at the right of the root   in such way that the higher has the low label
and labels increase as we go down along the vertical line (ordering increasing label vertices ``from high to low'').
Then iterate  this procedure for the descendants of each first generation vertex (i.e. the {\it second generation} vertices)
$i_1,\dots i_{s_0}$ and so on... (see figure \ref{fig:1}).
\setlength{\unitlength}{1cm}

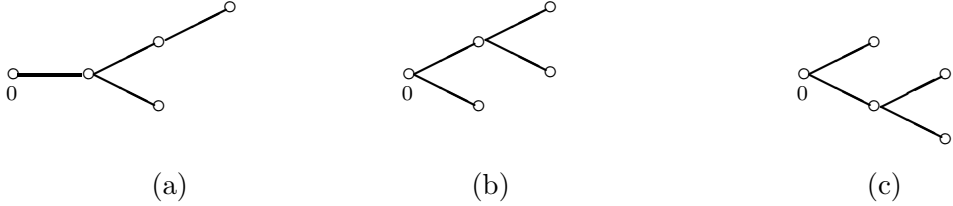
\begin{figure}[h]
\begin{center}
\begin{picture}(3,3)
\thicklines
\put(0,1.5){$\circ$}
\put(0,1.25){$\scriptstyle 0$}
\put(0.15,1.58){\line(1,0){0,85}}
\put(1,1.5){$\circ$}
\put(1.17,1.59){\line(2,1){0.8}}
\put(1.93,1.93){$\circ$}
\put(1.17,1.59){\line(2,-1){0.8}}
\put(1.93,1.07){$\circ$}
\put(2.12,2.05){\line(2,1){0.8}}
\put(2.88,2.39){$\circ$}
\put(1.93,0){(a)}
\end{picture}
\hspace{2cm}
\begin{picture}(3,3)
\thicklines
\put(0,1.5){$\circ$}
\put(0,1.25){$\scriptstyle 0$}
\put(0.17,1.59){\line(2,1){0.8}}
\put(0.93,1.93){$\circ$}
\put(0.17,1.59){\line(2,-1){0.8}}
\put(0.93,1.07){$\circ$}
\put(1.12,2.05){\line(2,1){0.8}}
\put(1.88,2.39){$\circ$}
\put(1.12,2.05){\line(2,-1){0.8}}
\put(1.88,1.53){$\circ$}
\put(0.93,0){(b)}
\end{picture}
\hspace{2cm}
\begin{picture}(3,3)
\thicklines
\put(0,1.5){$\circ$}
\put(0,1.25){$\scriptstyle 0$}
\put(0.17,1.59){\line(2,1){0.8}}
\put(0.93,1.93){$\circ$}
\put(0.17,1.59){\line(2,-1){0.8}}
\put(0.93,1.07){$\circ$}
\put(1.12,1.16){\line(2,1){0.8}}
\put(1.88,1.5){$\circ$}
\put(1.12,1.16){\line(2,-1){0.8}}
\put(1.88,0.64){$\circ$}
\put(0.93,0){(c)}
\end{picture}
\end{center}
\caption{\footnotesize{
the planar rooted trees  associated to the trees  ${\rm (a)}$ with edge set
$\{0,3\},\{1,3\}, \{2,3\}, \{1,4\}$,  ${\rm (b)}$ with edge set $\{0,2\},\{0,3\}, \{1,2\}, \{2,4\}$ and ${\rm (c)}$ with edge set
$\{0,2\},\{0,4\}, \{4,3\}, \{1,4\}$. Observe that ${\rm (b)}$ and ${\rm (c)}$ are {\it different} planar rooted tree
}}
\label{fig:1}
\end{figure}

There is a  natural { partial order} $ \prec$ among the
vertices in a rooted tree.
For $u,v\in t$, we say that { $u$ precedes   $v$} and write $u\prec v$ (or $v\succ u$)
if the (unique) path from the root  to $v$  contains $u$. If $\{v, u\}$
is an edge of $t$ rooted tree, then either $v\prec u$ or $u\prec v$. If $u\prec v$.
$u$ is called the {\it predecessor} and $v$ is called the {\it descendant}.
The root has no predecessor and it is the extremum  respect to the partial order relation $\prec$ in $t$.
For each  vertex $v$ of $t$, we will denote by { $s_v$}  the   {\it branching factor} of $v$ and we denote by
{ $v^1\,\dots,v^{s_v}$} the
 $s_v$ descendants of $v$,
($v^1$ being the higher and $v^{s_v}$ being the lower  in the drawing).

Clearly the  map $\t\mapsto m(\t)=t$ is many-to-one and the cardinality of the
preimage of a planar rooted tree
$t$ (=number of ways of labelling the $n$ non-root vertices of the tree with $n$ distinct labels
consistently with the rule ``from high to low") is given by
$$
\card{\{\t\in T^0_n: m(\t)=t\}}\;=\; {n!\over\prod_{v\succeq 0} s_{v_i}!}\Eq(rel1)
$$
We denote by $ {\TT}^0_n=$  the set of all planar rooted trees
with $n$ vertices and by  $\trp^{0,k}$ the
set of planar rooted trees with maximal generation number $k$; let also $\TT^0= \cup_{n\ge 0}{\TT}^0_n=\cup_{k\ge 0}\trp^{0,k}$ be
the set of all planar rooted trees.

Let now  $\m: \PP\to [0,\infty)^\PP: \g\mapsto \m_\g$ be
a positive valued function defined in $\PP$ and let, for each $\g\in \PP$, $R_\g \in[0,\infty)^\PP$ be defined by the equations
$$
\m_\g= R_\g \varphi_{\g}(\mu),\;\;\;\;\;\;\g\in \PP\Eq(muR)
$$

\\with
$$
\varphi _{\g} (\mu)\;=\; 1+\sum_{n\geq 1}
\sum_{(\g_{1},\dots ,\g_{n})\in\PP^n} b_n(\g;\g_1,\ldots,\g_n)\,
{\mu_{\g_1}}\dots{\mu_{\g_n}}
\Eq(r.7.1)
$$
for certain functions $b_n:\PP^{n+1}\to [0,\infty)$.
Denoting  $R_\g \varphi_{\g}(\mu)=T_{\g_0}(\mu)$ the equation \equ(muR) can be visualized
 in the diagrammatic form

$$
\begin{picture}(14,2.5)
\thicklines
\put(0.5,1.5){$\bullet$}
\put(0.5,1.3){$\scriptstyle\g_0$}
\put(1,1,5){$\doteq$}
\put(1.5,1.5){$\mu_{\g_0}$}
\put(2.1,1.5){=}
\put(2.6,1.5){{$T_{\g_0}(\mu)$}}
\put(4,1.5){$\doteq$}
\put(4.5,1.5){$\circ$}
\put(4.5,1.3){$\scriptstyle\g_0$}
\put(5,1.5){$+$}
\put(5.7,1.5){$\circ$}
\put(5.7,1.3){$\scriptstyle\g_0$}
\put(5.85,1.58){\line(1,0){1}}
\put(6.8,1.5){$\bullet$}
\put(6.8,1.3){$\scriptstyle\g_1$}
\put(7.2,1.5){$+$}
\put(7.7,1.5){$\circ$}
\put(7.7,1.3){$\scriptstyle\g_0$}
\put(7.85,1.58){\line(2,1){0.8}}
\put(8.65,1.89){$\bullet{\scriptstyle \g_1}$}
\put(7.85,1.58){\line(2,-1){0.8}}
\put(8.65,1.08){$\bullet{\scriptstyle \g_2}$}
\put(9.1,1.5){$+$}
\put(9.6,1.5){$\cdots$}
\put(10.3,1.5){$+$}
\put(10.8,1.5){$\circ$}
\put(10.8,1.3){$\scriptstyle\g_0$}
\put(10.97,1.58){\line(1,1){0.7}}
\put(11.64,2.23){$\bullet{\scriptstyle \g_1}$}
\put(10.97,1.58){\line(2,1){0.8}}
\put(11.64,1.86){$\bullet{\scriptstyle \g_2}$}
\put(11.64,1.3){\vdots}
\put(10.97,1.58){\line(1,-1){0.7}}
\put(11.64,0.74){$\bullet{\scriptstyle \g_n}$}
\put(12.09,1.5){$+$}
\put(12.57,1.5){$\cdots$}
\end{picture}
$$
The sum is over all single-generation rooted trees.  In each tree,
vertices with open circles with subscript $\g$ represents a factor $R_\g$, vertices with bullets with
subscript $\g$ a factor $\mu_\g$ and
vertices other than the root must be summed over all possible polymers
$\g$.  At each vertex with $n$ descendants, a ``vertex function'' $b_n$
acts, having as arguments the  $n+1$-tuple formed by the polymer
at the vertex and the $n$ polymers associated to the $n$ descendants of that vertex.
With this representation, the iteration $T^2(\mu)=T(T(\mu))$ corresponds to
replacing each of the bullets by each one of the diagrams of the
expansion for $T$.  This leads to planar rooted trees of up to two
generations, with open circles at first-generation vertices and bullets
at second-generation ones.  In particular, all single-generation trees
have only open circles.  Notice that the two drawings of Figure
\ref{fig:1} appear in two different terms of the expansion, and hence
should be counted as \emph{different} diagrams.  More generally, the
$k$-th iteration of $T$ involves all possible planar rooted trees up to $k$
generations.
In each term of the expansion, vertices of generation $k$ are occupied
by bullets and all the others by open circles.
A straightforward inductive argument shows that
$$
T^k_{\g_0}(\mu) \;=\; R_{\g_0} \Bigl[\sum_{\ell=0}^{k-1}
\Phi^{(\ell)}_{\g_0}( R) + \Phi^{(k)}_{\g_0}( R,\mu)\Bigr]
\Eq(r.8.1)
$$
where we have denoted $R=\{R_\g\}_{\g\in \PP}$ and
$$
\Phi^{(\ell)}_{\g_0}(R) \;= \,\sum_{t \in\trp^{0,\ell}}\,
\prod_{v\succeq 0}
\Bigg\{\sum_{(\g_{v^1},\dots,\g_{v^{s_v}})\in \PP^{s_v}}b_{s_v}(\g_{v};\g_{v^1},\dots,\g_{v^{s_v}})\,
R_{\g_{v^1}}\dots R_{\g_{v^{s_v}}}
\Bigg\}\Eq(explw3)
$$
Here the product $\prod_{v\succeq 0}$ over the vertices of $t$ must be done respecting the partial order of the set of vertices in $t$,
i.e. if $v\succ u$ the $v$ must be at the right of $u$ in the product.
The factor  $\Phi^{(k)}_{\g_0}(R,\mu)$ has a similar expression but
with the activities of the vertex of the $k$-th generation
weighted by $\mu$.
Here we agree that $b_0(\g_v)\equiv 1$
and $\prod_\emptyset \equiv 1$.
We are interested in the $\ell\to\infty$ limit
of \equ(explw3).

\begin{proposition}\label{prop:1}
Let  $\m: \PP\to [0,\infty)^\PP: \g\mapsto \m_\g$ be a positive valued function and
let, for each $\g\in \PP$, $R_\g \in[0,\infty)^\PP$ be defined by the equations
\equ(muR). Let, $\forall\g\in\PP$, $\tilde\r_\g\in[0,\infty)$ such that $\tilde\r_\g\le R_\g$.
Then the series
$$
\Phi_{\g_0} ({\tilde\r})\;\bydef\;
\,\sum_{t \in\trp^0}
\prod_{v\succeq 0}
\Bigg\{\sum_{(\g_{v^1},\dots,\g_{v^{s_v}})\in \PP^{s_v}}b_{s_v}(\g_{v};\g_{v^1},\dots,\g_{v^{s_v}})\,
\tilde\r_{\g_{v^1}}\dots \tilde\r_{\g_{v^{s_v}}}
\Bigg\}
\Eq(r.11)
$$
is finite for each $\g_0\in\PP$.  Furthermore
$$ \tilde\r_{\g_0}\Phi_{\g_0}({\tilde\r})\le \mu_{\g_0}
\Eq(r.12)
$$
for each $\g_0\in\PP$.
\end{proposition}

\proof    By definition
$\Phi_{\g_0}(\tilde\r)=  \sum_{\ell=0}^{\infty} \Phi^{(\ell)}_{\g_0}(\tilde\r)$.
By \equ(r.8.1), the fact that $T^k_{\g_0}(\mu)=\m_{\g_0}$ for all $k\in \mathbb{N}$, and the assumption $\tilde\r_\g\le R_\g$
for all $\g\in \PP$, we obtain that
$$
\tilde\r_{\g_0}\sum_{\ell=0}^{n}\Phi^{(\ell)}_{\g_0}(\tilde\r)\;\le\; R_{\g_0}\sum_{\ell=0}^{n}\Phi^{(\ell)}_{\g_0}( R)\;\le\;
\m_{\g_0}
$$

\\for all $n$. Thus, since the sequence of partial sums of the series $\r_{\g_0}\Phi_{\g_0}(\tilde \r)$ is monotonic increasing
and bounded by $\m_{\g_0}$,
$\tilde\r_{\g_0}\Phi_{\g_0}(\tilde \r)$
converges, and  $\tilde\r_{\g_0}\Phi_{\g_0}(\tilde \r)\le \mu_{\g_0}$ .
$\qed$

\vv
\vv

\\{\bf 3.3. End of the proof of theorem 1}
\vv
\\We first reorganize the sum over labelled trees appearing
in formula \equ (mukotpre) in terms of the called planar rooted trees previously introduced.
Namely, recalling that
$ {\TT}^0_n$  is the set of all planar rooted trees
with fixed root $0$ and $n$ vertices (different from the root), we can rewrite the r.h.s. of \equ(mukotpre) as
$$
|\tilde\Pi|^{\g_0}_{\PP}(\tilde\r)\,= \, 1+
\sum_{n=1}^\io{1\over n!}\sum\limits_{t\in \TT^0_{n}}\sum_{\t\in T^0_ {n}\atop m(\t)= t}
\sum_{(\g_{1},\dots ,\g_{n})\subset\PP^n}
\prod_{\{i,j\}\in E_\t}  F(\g_i,\g_j){\tilde\r_{\g_1}}\dots{\tilde\r_{\g_n}}\Eq(aa)
$$
Observe that the factor
$$
\sum_{(\g_{1},\dots ,\g_{n})\subset\PP^n}
\prod_{\{i,j\}\in E_\t}  F(\g_i,\g_j){\tilde\r_{\g_1}}\dots{\tilde\r_{\g_n}}
$$
depends only on the  planar rooted tree $t=m(\t)$ associated to $\t$ (labels
 of $\t$ are dummy indices in the sum), i.e.
$$
\sum_{(\g_{1},\dots ,\g_{n})\subset\PP^n}
\prod_{\{i,j\}\in E_\t}  F(\g_i,\g_j){\tilde\r_{\g_1}}\dots{\tilde\r_{\g_n}}
=
\prod_{v\succeq v_0}
\Bigg\{\prod_{i=1}^{s_v}\sum_{\g_{v^i}\in \PP}
F(\g_v,\g_{v^i})\tilde\r_{\g_{v^i}}
\Bigg\}\Eq(explw1)
$$
with the convention that $\prod_{i=1}^{s_v}
\sum_{\g_{v^i}\in\PP}F(\g_v,\g_{v^i})\tilde\r_{\g_{v^i}}=1$ when $s_v=0$.

So in conclusion, inserting \equ(explw1) into \equ(aa) and using also \equ(rel1), we obtain
$$
|\tilde\Pi|^{\g_0}_{\PP}(\tilde\r)\,=
\sum\limits_{t\in \TT^0}
\prod_{v\succeq v_0}
{1\over s_v!}
\Bigg\{\prod_{i=1}^{s_v}\sum_{\g_{v^i}\in \PP}
F(\g_v,\g_{v^i})\tilde\r_{\g_{v^i}}
\Bigg\}
\Eq(aab)
$$

\\Comparing \equ(aab) with \equ(r.11)  we immediately
see that
$|\tilde\Pi|^{\g_0}_{\PP}(\tilde\r)=\Phi_{\g_0}(\tilde\r)$
provided
$$
b_n(\g;\g_1,\ldots,\g_n)= {1\over n!}\prod_{i=1}^{n}F(\g,\g_i)\Eq(cig)
$$
so that
$$
\varphi _{\g} (\mu)\;=\; 1+\sum_{n\geq 1} \frac{1}{n!}\,
\sum_{(\g_{1},\dots ,\g_{n})\in\PP^n}\prod_{i=1}^{n}F(\g,\g_i)
{\mu_{\g_i}}\;=\; e^{\sum_{\gt\in\PP}F(\g,\gt)\mu_{\g}}
\Eq(r.15.2)
$$

Hence proposition \ref{prop:1} yields
the criterion \equ(muRv) for the convergence of the series $|\tilde\Pi|^{\g_0}_{\PP}(\tilde\r)$ defined in  \equ(TP).
As a matter of fact,
by proposition \ref{prop:1}, with the identification
\equ(cig), we have immediately that the series $|\tilde\Pi|^{\g_0}_{\PP}(\tilde\r)$ defined in  \equ(mukotpre)
is finite for each $\g_0\in\PP$ and
$\tilde\r_{\g_0}
|\tilde\Pi|^{\g_0}_{\PP}(\tilde\r)\le \mu_{\g_0}$
for each $\g_0\in\PP$.
Now recalling \equ(dises) and \equ(rhotil) we obtain $\r_{\g_0}\Pi_{\g_0}(\r)\le \m_{\g_0}$.
\vv

\vv\vv\def\LL{{\mathbb{L}}}
\numsec=4\numfor=1
\\{\Large\bf 4. Example.  BEG model with infinite range interactions in the low temperature disordered phase
}
\vv

As an example, we consider the Blume-Emery-Griffiths (BEG)  model \cite{BEG}
with infinite range interactions in the low temperature disordered
phase. The model is defined on the cubic unit lattice in
$d$-dimensions $\Z^d$ by supposing that in each vertex $x\in \mathbb{Z}^d$
there is a spin variable $\s_x$ taking values in the set
$\{0,-1,+1\}$. These spins interact via the (formal) Hamiltonian

$$
H=-\sum_{\{x,y\}\subset \Z^d}[J_{xy}\s_x\s_y+ K_{xy}\s^2_x\s^2_y]+ D\sum_{x\in \Z^d}\s_x^2
\Eq(Ham)
$$
where  $J_{xy}\ge 0$ and $K_{xy}\in \mathbb{R}$  are summable interactions and we put
$$
J={1\over 2}\sup_{x\in \mathbb{Z}^d}\sum_{y\in \mathbb{Z}^d\atop y\neq x}(J_{xy}+|K_{xy}|)\Eq(defJ)
$$
In the region of parameters
$$
D>J\Eq(dis)
$$
the ground state is
$\s=0$. This
region is called the disordered phase. If
$J_{xy}$ and $K_{xy}$ are  nearest neighbor interactions
(or  finite range), the low temperature disordered phase
can be studied using the standard Pirogov-Sinai theory.

We  will make here different assumptions on the interactions $J_{xy}$ and $K_{xy}$.
Namely, we  suppose that there exist
positive constants  $c$, $J_1$, $\l$ and $\l'$ (with $0<\l<\l'$)
  such that

$$
J_{xy}+|K_{xy}|\le {2J_1\over |x-y|^{d + \l} }~~~~~~~~\forall\{x,y\}\in \Z^d\Eq(sum)
$$
and
$$
J_{xy}\ge {c\over |x-y|^{d + \l'} }\,\,\,\,\,\,{\rm or}\,\,\,\,\,\, |K_{xy}|\ge {c\over |x-y|^{d + \l'} }\Eq(sum2)
$$
where $|x-y|$ is the usual nearest neighbor path distance, i.e., $|x-y|$ is the length of the
shortest path of nearest neighbors connecting $x$ to $y$.
Due to the assumption \equ(sum2) the low temperature  phase of the BEG  model described by the Hamiltonian
\equ(Ham), cannot be studied using the standard low temperature Pirogov-Sinai, which explicitly requires finite range
interactions. If we further assume  that the polynomial decay is  slow, e.g. by supposing
$$
\l'<2d+1 \Eq(park)
$$
then this model is not even  included in the class of  models
whose low temperature phase can be studied
via the extension of the Pirogov-Sinai theory to infinite range interactions given in \cite{Pa}.

We'll show in this section  that the partition function of the spin model described by Hamiltonian \equ(Ham)
can rewritten as the partition function of a polymer system of the type considered in the previous sections.
Then, using theorem 1, we will prove that, in the disordered phase \equ(dis)  and with the assumptions
\equ(sum), \equ(sum2), \equ(park), such polymer expansion converges for sufficiently low temperatures.

In order to do that, let us put the system in a finite box  $\L\subset \Z^d$ and let us  define,
for a fixed spin configuration $\s_\L$ in $\L$,
the subset of $\L$ given
by  $P=\{x\in \L: \s_x\neq 0\}$. We view this set as the union of its  connected components,
i.e. $P=\cup_{i=1}^n p_i$ with each set $p_i\subset \L$  being connected in the sense that
for each partition $A,B$ of $p_i$ (i.e. $A\cup B=p_i$ and $A\cap B=\emptyset$) there exist $x\in A$ and $y\in B$ such that
$|x-y|=1$.
The configuration $\s_\L$ induces a (non zero) spin configuration $s_{p_i}$ on each connected component
$p_i$ of $P$ which is a function
$s_{p_i}:p_i\to \{-1,+1\}: x\mapsto s_x$. The pairs ${\bm p}_i=(p_i,s_{p_i})$ are the polymers associated to the configuration $\s_\L$.
By construction the correspondence $\s_\L \leftrightarrow \{\bm p_1,\dots,\bm p_n\}$ is one to one.
The distance between two polymers $\bm p=(p,s_p)$ and $\tilde{\bm p}=(\tilde p,s_{\tilde p})$
 is the number $d(p,\tilde p)=\min_{x\in p,\,\,y\in \tilde p}|x-y|$. Note that if $ \{\bm p_1,\dots,\bm p_n\}$ are the polymers
associated to the configuration $\s_\L$, then necessarily $d(p_i,p_j)\ge 2$ for all $\{i,j\}\subset \{1,\dots, n\}$.

With these definitions we can rewrite the Hamiltonian of the system
in a  box $\L\subset \Z^d$ with free boundary conditions as (here below $\b$ is the inverse temperature)
$$
\b H_\L(\s)=\sum_{1\le i<j\le n }W(\bm p_i,\bm p_j)+ \sum_{i=1}^n
\Big[\b D|p_i|-A(\bm p_i)\Big]
$$
where
$$
W(\bm p_i,\bm p_j)=-\b\sum_{x\in p_i\atop y\in p_j} [J_{xy}s_x s_y+K_{xy}]\Eq(longr)
$$
$$
A(\bm p_i)=\b\sum_{\{x,y\}\subset p_i}  [J_{xy}s_xs_y+K_{xy}]\Eq(Ap)
$$
Observe now that to sum over configuration $\s_\L$ in $\L$ is equivalent to sum over polymers configurations
$\{\bm p_1,\dots,\bm p_n\}$ in $\L$ such that $n\ge 0$ ($n=0$, i.e. no polymers, is the ground state configuration)
and $d(p_i,p_j)\ge 2$ for all pairs $\{i,j\}\subset \{1,\dots,n\}$.
Hence the partition function of the system, at inverse temperature $\b$ and with free boundary conditions, is rewritten as
$$
Z_\L(\b)=\sum_{\s_\L}e^{-\b H (\s_\L)}~~~~~~~~~~~~~~~~~~~~~~~~~~~~~~~~~~
~~~~~~~~~~~~~~~~~~\Eq(parttt)
$$
$$
=1+\sum_{n\ge 1}{1\over n!}\sum_{(\bm p_1,\dots,\bm p_n)\in \PP_\L^n \atop d(p_i,p_j)\ge 2}\r_{\bm p_1}\dots\r_{\bm p_n}
e^{-\sum_{i<j}W(\bm p_i,\bm p_j)}\Eq(parttt2)
$$
where
$$
\r_{\bm p}=e^{-\big[\b D|p|-A(\bm p)\big]}\Eq(actp)
$$
and
$$
\PP_\L=\{\bm p=(p,s_p): \mbox{$p\subset \L$ connected, $s_p$ function from $p$ to $\{-1,+1\}$}\}
$$

\\We now extend the definition of $W(\bm p_i,\bm p_j)$ to all pairs in $\PP$ as
$$
W(\bm p_i,\bm p_j)=\cases{-\b\sum\limits_{x\in p_i\atop y\in p_j} [J_{xy}s_x s_y+K_{xy}]~~~~~~~~ &if $d(p_i,p_j)\ge 2$\cr\cr
+\infty & otherwise}
\Eq(extlongr)
$$

With these definitions it is immediate to see that r.h.s. of \equ(parttt2) can be written as
$$
Z_\L(\b)= 1+\sum_{n\ge 1}{1\over n!}
\sum_{(\bm p_1,\dots,\bm p_n)\in \PP^n_\L}
\r_{\bm p_1}\dots\r_{\bm p_n}
e^{-\sum_{1\le i<j\le n}W(\bm p_i,\bm p_j)}\Eq(parttt3)
$$
which is the partition function of a polymer gas of the type \equ(partiz) in which the polymers
are  elements of the set
$\PP$ defined by
$$
\PP=\Big\{\bm p=(p,s_p): \mbox{$p\subset \Z^d$ connected and finite,
$s_p$ function from $p$ to $\{-1,+1\}$} \Big\}  \Eq(polsp)
$$
with activity  given in \equ(actp) and with incompatibility relation $\bm p \not\sim\tilde{\bm p}\Leftrightarrow
d(p,\tilde p)< 2$.
This pair interaction $W(\bm p_i,\bm p_j)$  is stable in the sense of \equ(stab).
As a matter of fact it is easy to check that, for all $n\in \mathbb{N}$ and all
$(\g_1, \dots, \g_n)\in \PP^n$.
$$
\sum_{1\le i<j\le n}W(\bm p_i,\bm p_j)\ge -\sum_{i=1}^{n}B(\bm p_i)
$$
with
$$
B(\bm p_i)= \b J|p_i|-A(\bm p_i)
$$
where $J$ is defined in \equ(defJ) and $A(\bm p_i)$ is defined in \equ(Ap).
Again note that we have to check this condition on non intersecting sets of polymers
since when $(\g_1, \dots, \g_n)$ contains one o more incompatible pairs this inequality is trivially satisfied.

So, by theorem 1,  the pressure of this polymer gas (i.e. the free energy of our long range BEG model) is absolutely convergent
if there exist $\m_{\bm p}$ such that
such that
$$
e^{-\b (D-J)|p|}\le {\m_{\bm p}}\,e^{-\sum_{\tilde{\bm p}\in\PP} F(\bm p,\tilde{\bm p})
{\mu_{\tilde{\bm p}}}}, \;\;\;\,\,\,\,\,\,\,\forall \g\in \PP \Eq(muRv2)
$$
\def\pp{{\bm p}}\def\pt{{\tilde{\bm p}}}
We choose
$$
\m_\pp=e^{-\b (D-J)|p|} e^{\a|p|}\def\pp{{\bm p}} \Eq(mpp)
$$
Hence, inserting \equ(mpp) in \equ(muRv2), we obtain that
the pressure of such contour gas can be written in terms of an absolutely convergent series if, for some $\a>0$
$$
\sum_{\pt\in \PP}F(\pp,\pt)\,\m_\pt\le \a|p|\Eq(converg2)
$$
By bounding again $F(\pp,\pt)\le 1$ whenever $\pt\nsim\pp$ (recall that the short range potential $U$ is in this case purely hard core),
we get
$$
\sum_{\pt\in\PP} F(\pp,\tilde \pp)\,\m_\pt=
\sum_{\pt\in\PP \atop d(p,\tilde p)\le 1}
\m_\pt + \sum_{\pt\in\PP \atop  d(p,\tilde p)>1 }
|W(\pp,\tilde \pp)|\;\m_\pt\le
$$
$$
\le ~|p|\Big[2d\sup_{x\in \mathbb{Z}^d}
\sum_{\pt\in\PP \atop x\in \pp}\mu_\pt\Big]+
\max_{x\in  p}\sum_{\pt\in \PP \atop d(x,\tilde p)>1}
|W(\pp,\tilde \pp)|\;\mu_\pt
$$
where $d(x,\tilde p)= \min_{\,y\in \tilde p}|x-y|$.
Observe now that,  by  \equ(sum),
$|W(\pp,\tilde \pp)|\le \b J_1|p||\tilde p|n^{-(d+\l)}$ whenever $d(p,\tilde p)=n$.
Therefore
$$
\max_{x\in p}\sum_{\pt\in \PP \atop d(\pt,x)>1}
|W(\pp,\tilde \pp)|\; \mu_\pt\le
|p|\sum_{n>1}{\b J_1\over n^{d+\l}}\max_{x\in  p}
\sum_{\pt\in\PP\atop d(\tilde p,x)=n}|\tilde p|\; \mu_\pt \le
$$
$$
\le
|p|\sum_{n>1}{\b J_1\over n^{d+\l}}
\sup_{x\in \mathbb{Z}^d}\sum_{\pt\in\PP\atop \tilde p\cap S_n(x)\neq \emptyset}|\tilde p|\; \mu_\pt
\le |p|\sum_{n>1}{\b J_1\over n^{d+\l}}|S_n|
\sup_{x\in \mathbb{Z}^d}\sum_{\pt\in\PP\atop x\in \tilde p}|\tilde p|\; \mu_\pt\le
$$
where $S_n=\{y\in \Z^d: |y|=n\}$.
An easy calculation show that
$$
|S_n|
\le {(2d)^d\over d!}n^{d-1}
$$
So we get
$$
\max_{x\in p}\sum_{\pt\in \PP \atop d(\pt,x)>1}
 |W(\pp,\tilde \pp)| \;\mu_\pt\le
\b J_2|p|
\sup_{x\in \mathbb{Z}^d}\sum_{\pt\in\PP\atop x\in \tilde p}|\tilde p| \;\m_\pt
$$
where
$$
J_2={(2d)^d J_1\over d!}  \sum_{n=2}^\infty\,\,{1\over n^{1+\l}}
$$

\\Hence
$$
\sum_{\pt\in\PP} F(\pp,\tilde \pp)\;\m_\pt
\le |p|\Bigg[\Big(2d\sup_{x\in \mathbb{Z}^d}
\sum_{\pt\in\PP \atop x\in \pp}\;\m_\pt\Big)+
\b J_2 \sup_{x\in \mathbb{Z}^d}
\sum_{\pt\in\PP\atop x\in \pp}|\tilde p |\;\m_\pt\Bigg]\le
$$
$$
\le  |p|\Big[2d +
\b  J_2 \Big]\sup_{x\in \mathbb{Z}^d}
\sum_{\pt\in\PP\atop x\in \pt}|\tilde p|\;\m_\pt
\le  J_\b | p|
\sum_{\pt\in\PP\atop x\in \pt}|\tilde p|\;\m_\pt
$$

\\where
$$
J_\b= 2d + \b J_2
$$


\\Therfore, recalling  \equ(mpp),
convergence condition \equ(converg2) becomes
$$
J_\b\sum_{n=1}^\infty n [e^{-\b (D-J)} e^{\a}]^{n}2^nC_n\le \a\Eq(quasi)
$$
where $C_n$ is the number of connected sets of vertices of $\Z^d$ with cardinality $n$
containing the origin (the factor $2^n$ in l.h.s. of \equ(quasi) counts the number of functions
from $p$ to $\{-1,+1\}$ when $|p|=n$). $C_n $ can be easily bounded by $C^n$ for some $C$, e.g. one can
take $C_n\le (4d)^n$. So condition \equ(quasi) becomes
$$
\sum_{n=1}^\infty n (xe^\a)^{n}\le {\a\over J_\b}\Eq(conde)
$$
where
$$
x= 8d e^{-\b (D-J)} \Eq(defx)
$$
Formulas \equ(conde) and \equ(defx)  imply that
$$
e^{-\b (D-J)}\le \Big[e^{-\a}f(\a/J_\b)\Big]{1\over 8d }
$$
where
$$
f(u)={2u\over2u+1+\sqrt{4u+1}}
$$
For example, taking  $\a=1/2$ and bounding $f(u)\le 2u/(2u+1)$
(we are not looking here for optimal estimates), we obtain that
convergence occurs if
$$
e^{-\b (D-J)}\le  {1\over 8d\sqrt{e}(2d+1+\b J_2)}
$$
i.e.,  for all inverse temperatures $\b\ge \b_0$,
where $\b_0$ is the positive solution of the equation
$$
(2d+1+\b J_2)= {e^{\b (D-J)}\over 8\sqrt{e}\,d}
$$

\vglue.5truecm
\\{\bf Acknowledgements}.

\\This work was supported by CNPq (Conselho Nacional de
Desenvolvimento Cient\'{\i}fico e Tecnol\'ogico), a Brazilian
Governmental agency promoting scientific and technology
development.

\def\0{\emptyset}

\end{document}